# Entropy and Information of a harmonic oscillator in a time-varying electric field in 2D and 3D noncommutative spaces

J. P. G. Nascimento*, V. Aguiar and I. Guedes

Departamento de Física, Universidade Federal do Ceará, Campus do PICI, Caixa Postal 6030, 60455-760, Fortaleza, CE, Brazil.

**Abstract**

We analyzed the noncommutativity effects on the Fisher information ($F_{\hat{r},\hat{p}}$) and Shannon entropies ($S_{\hat{r},\hat{p}}$) of a harmonic oscillator immersed in a time-varying electric field in two and three dimensions. We find the exact solutions of the respective time-dependent Schrödinger equation and use them to calculate the Fisher information and the Shannon entropy for the simplest case corresponding the lowest-lying state of each system. While there is no problem in defining the Shannon entropy for noncommutating spaces, the definition of the Fisher information had to be modified to satisfy the Cramer-Rao inequalities. We observe for both systems, how the Fisher information and Shannon entropy in position and momentum change due to the noncommutativity of the space. We verified that the Bialynicki-Birula-Mycielski (BBM) entropic uncertainty relation still holds in the systems considered.

*Corresponding author. Tel.: +55 85999109489.

E-mail address: joaopedro@fisica.ufc.br (J. P. G. Nascimento).

# 1. Introduction

Since it was introduced in literature as a formalism to study string and field theories, the noncommutativity of spacetimes has been considered for several authors [1-10]. Most of them, investigate either classical or quantum mechanics of a particle confined by a quadratic potential in the generalized time-independent noncommutative backgrounds.

Investigations of the quantum motion of harmonic oscillators in the presence of magnetic and electric fields in noncommutative backgrounds have also been reported [11-17]. In Ref. [12], Liang and Yang studied the motion of a time-dependent harmonic oscillator in a magnetic and an electric field in a time-independent 3D noncommutative background. They obtained the wave functions, geometric phases and explicit forms of the coherent states which are not minimum uncertainty states for the coordinates and momenta. They showed that the oscillator feels an effective magnetic field along a new direction owing to the momentum–momentum noncommutativity.

In Ref. [15], Dey and Fring investigated a time-independent harmonic oscillator on a time-dependent 2D noncommutative space. They used the Lewis and Riesenfeld method of invariants [18] to construct explicit analytical solutions for the corresponding time-dependent Schrödinger equation. They used those solutions to to verify a generalized version of Heisenberg´s uncertainty relations for which the lower bound becomes a time-dependent function of the background space.

Besides the Heisenberg uncertainty relation, which imposes a lower limit for the product of uncertainties in position and momentum, other uncertainty relations beyond that based on standard deviations, and based on the Shannon entropy [19] and Fisher information [20] have been considered [21-27].

We are aware of no paper reporting the calculation of either the Fisher information or Shannon entropy in quantum systems on noncommutative backgrounds. Therefore, here we investigate the quantum harmonic oscillator in time-varying electric field in the framework of noncommutative 2D and 3D backgrounds. This paper is outlined as follows. In Sections 2 and 3, we calculate the wave functions, the Fisher information and the Shannon entropy for a time-independent harmonic oscillator in the presence of time-dependent electric field in both 2D and 3D time-independent backgrounds. In Section 4, we present the concluding remarks.

## 2. Harmonic oscillator in a time-dependent electric field on a 2D noncommutative phase space.

Consider a noncommutative phase space (NCPS), where both the spatial–spatial and momentum-momentum coordinates do not commute. By denoting the position and momentum operators in the noncommutative Quantum Mechanics as $\hat{r}$ and $\hat{p}$, respectively, we obtain the following relations for a 2D problem [6]

$$[\hat{x}_m, \hat{x}_n] = i\theta_{mn}, \qquad [\hat{p}_m, \hat{p}_n] = i\eta_{mn}, \qquad [\hat{x}_m, \hat{p}_n] = i\hbar\left(\delta_{mn} + \frac{1}{4\hbar^2}\theta_{ml}\eta_{nl}\right), \quad (1)$$

with $m, n = 1, 2$, where $\theta_{mn}$ and $\eta_{mn}$ are real antisymmetric tensors associated with the noncommutativity of the operators. By using the relations

$$\hat{x}_m = x_m - \frac{1}{2\hbar}\theta_{mn}p_n, \tag{2.a}$$

$$\hat{p}_m = p_m + \frac{1}{2\hbar}\eta_{mn}x_n, \tag{2.b}$$

we can map the noncommutative space into the commutative one and obtain

$$[x_m, x_n] = 0, \qquad [p_m, p_n] = 0, \qquad [x_m, p_n] = i\hbar\delta_{mn}, \tag{3}$$

*i.e.*, the operators $r$ and $p$ in the commutative space satisfy the usual Heisenberg uncertainty relations. By considering $\theta_{mn} = \epsilon^{mn}\theta$ and $\eta_{mn} = \epsilon^{mn}\eta$, we obtain

$$\hat{x}_1 = x_1 - \frac{1}{2\hbar}\theta p_2, \qquad \hat{x}_2 = x_2 + \frac{1}{2\hbar}\theta p_1, \tag{4.a}$$

$$\hat{p}_1 = p_1 + \frac{1}{2\hbar}\eta x_2, \qquad \hat{p}_2 = p_2 - \frac{1}{2\hbar}\eta x_1. \tag{4.b}$$

To calculate the Fisher Information and Shannon Entropy for a harmonic oscillator in a time-dependent electric field in a 2D noncommutative background, we have to solve the time-dependent Schrödinger equation

$$i\hbar\partial_t\psi^{NCPS}(\hat{r},t) = H^{NCPS}(\hat{r},\hat{p},t)\psi^{NCPS}(\hat{r},t) \tag{5}$$

where $H^{NCPS}$ is given by

$$H^{NCPS}(\hat{r},\hat{p},t) = \frac{1}{2m}(\hat{p}_1^2 + \hat{p}_2^2) + \frac{1}{2}m\omega_0^2(\hat{x}_1^2 + \hat{x}_2^2) + q\Phi(\hat{r},t), \tag{6}$$

where $\Phi(\hat{r},t) = E_1(t)\hat{x}_1 + E_2(t)\hat{x}_2$ is the eletrostatic potential.

To write Eq.(6) in terms of the operators $x_i$ and $p_i$ we use the Bopp shift method [15, 28], which consists in replacing Eqs. (4) into Eq. (6). Thus, we obtain from Eq. (5)

$$i\hbar\partial_t\psi^{NCPS}(r,t) = \left[\frac{1}{2M}(p_1^2 + p_2^2) + \frac{1}{2}M\omega_1^2(x_1^2 + x_2^2) - \omega_2 L_3\right]\psi^{NCPS}(r,t)$$
$$+ [A(t)p_1 + B(t)p_2 + C(t)x_1 + D(t)x_2]\psi^{NCPS}(r,t), \tag{7}$$

where

$$M = m\left(1 + \frac{m^2\omega_0^2\theta^2}{4\hbar^2}\right)^{-1}, \tag{8.a}$$

$$\omega_1^2 = \omega_0^2\left(1 + \frac{m^2\omega_0^2\theta^2}{4\hbar^2}\right)\left(1 + \frac{\eta^2}{4\hbar^2 m^2\omega_0^2}\right), \tag{8.b}$$

$$\omega_2 = \frac{1}{2\hbar m}(\eta + \theta m^2\omega_0^2), \tag{8.c}$$

$$A(t) = \frac{q\theta}{2\hbar}E_2(t), \tag{8.d}$$

$$B(t) = -\frac{q\theta}{2\hbar}E_1(t), \tag{8.e}$$

$$C(t) = qE_1(t), \tag{8.f}$$

$$D(t) = qE_2(t), \tag{8.g}$$

$$L_3 = x_1 p_2 - x_2 p_1. \tag{8.h}$$

To find the exact expressions for $\psi^{NCPS}(r,t)$ we proceed as follows. First, consider the time-dependent canonical transformation whose generating function is

$$G(x_1, x_2, P_1, P_2, t) = (P_1\cos\phi + P_2\sin\phi)x_1 + (-P_1\sin\phi + P_2\cos\phi)x_2, \tag{9}$$

where $\phi = \phi(t) = \omega_2 t$.

From Eq. (9) we obtain the new canonical variables

$$p_1 = \frac{\partial G}{\partial x_1} = P_1 cos\phi + P_2 sin\phi, \qquad (10.a)$$

$$p_2 = \frac{\partial G}{\partial x_2} = -P_1 sin\phi + P_2 cos\phi, \qquad (10.b)$$

$$Q_1 = \frac{\partial G}{\partial P_1} = cos\phi x_1 - sin\phi x_2, \qquad (10.c)$$

$$Q_2 = \frac{\partial G}{\partial P_2} = sin\phi x_1 + cos\phi x_2, \qquad (10.d)$$

and the new Hamiltonian, $K(Q_1, Q_2, P_1, P_2, t) = H^{NCPS}(x_1, x_2, p_1, p_2, t) + \partial_t G$, reads

$$K(\boldsymbol{Q}, \boldsymbol{P}, t) = \frac{1}{2M} P_1^2 + + \frac{1}{2} M\omega_1^2 Q_1^2 + \Omega_1(t) P_1 + \xi_1(t) Q_1$$
$$+ \frac{1}{2M} P_2^2 + + \frac{1}{2} M\omega_1^2 Q_2^2 + \Omega_2(t) P_2 + \xi_2(t) Q_2, \qquad (11)$$

where

$$\Omega_1(t) = A(t)cos\phi - B(t)sin\phi, \quad \Omega_2(t) = A(t)sin\phi + B(t)cos\phi, \qquad (12.a)$$
$$\xi_1(t) = C(t)cos\phi - D(t)sin\phi, \quad \xi_2(t) = C(t)sin\phi + D(t)cos\phi. \qquad (12.b)$$

Equation (11) represents the sum of two uncoupled Hamiltonians of time-dependent quadratic harmonic oscillators [29-31]. The solution of the time-dependent Schrödinger equation for $K(\boldsymbol{Q}, \boldsymbol{P}, t)$ can be obtained by employing the Lewis and Riesenfeld method [18].

Consider the Schrödinger equation

$$i\hbar \partial_t \psi(Q, t) = H(Q, P, t)\psi(Q, t), \qquad (13)$$

where $H(Q, P, t)$ is given by

$$H(Q, P, t) = \beta_1 P^2 + \beta_2 Q^2 + \beta_3(t) P + \beta_4(t) Q, \qquad (14)$$

and $\beta_m$ ($m = 1, 2, 3, 4$) is the $m$-th coefficient of $H(Q, P, t)$.

The invariant $I$ associated to the Hamiltonian (14) is [32]

$$I(Q,P,t) = \frac{1}{4\rho^2(t)}\left(Q - Q_p^{cl}(t)\right)^2 + \left[\rho(t)\left(P - P_p^{cl}(t)\right) - \frac{\dot{\rho}(t)}{2\beta_1}\left(Q - Q_p^{cl}(t)\right)\right]^2, \quad (15)$$

where the dot corresponds to time derivative, $Q_p^{cl}(t)$ and $P_p^{cl}(t)$ are particular ($p$) solutions of the following classical ($cl$) equations of motion in coordinate and momentum spaces, respectively

$$\ddot{Q}_p^{cl}(t) - \frac{\dot{\beta}_1}{\beta_1}\dot{Q}_p^{cl}(t) + 4\beta_1\beta_2 Q_p^{cl}(t) = -\frac{\dot{\beta}_1\beta_3(t)}{\beta_1} - 2\beta_1\beta_4(t) + \dot{\beta}_3(t), \quad (16.a)$$

$$\ddot{P}_p^{cl}(t) - \frac{\dot{\beta}_2}{\beta_2}\dot{P}_p^{cl}(t) + 4\beta_1\beta_2 P_p^{cl}(t) = \frac{\dot{\beta}_2\beta_4(t)}{\beta_2} - 2\beta_2\beta_3(t) - \dot{\beta}_4(t), \quad (16.b)$$

and $\rho(t)$ satisfies the following auxiliary differential equation

$$\ddot{\rho}(t) - \frac{\dot{\beta}_1}{\beta_1}\dot{\rho}(t) + 4\beta_1\beta_2\rho(t) = \frac{\beta_1^2}{\rho^3(t)}. \quad (17)$$

From Eq. (14) we observe that the coefficients $\beta_1$ and $\beta_2$ for the two uncoupled Hamiltonians in Eq. (11) are time-independent. Therefore, the solution of Eq. (17) corresponds to the constant

$$\rho = \left(\frac{\beta_1}{4\beta_2}\right)^{1/4}. \quad (18)$$

According to Lewis and Riesenfeld [18], the solutions $\psi_n(Q,t)$ of the Schrödinger equation (13) are related to the eingenfunction ($\varphi_n(Q,t)$) of $I$ by

$$\psi_n(Q,t) = exp[i\Upsilon_n(t)]\varphi_n(Q,t), \quad (19)$$

where the phase functions $\Upsilon_n(t)$ satisfy the equation

$$\hbar\dot{\Upsilon}_n(t) = \langle\varphi_n(Q,t)|i\hbar\partial_t - H(Q,P,t)|\varphi_n(Q,t)\rangle. \quad (20)$$

The normalized eigenstates and the engenvelues of $I$ are given, respectively, by [32]

$$\varphi_n(Q,t) = \left(\frac{1}{2\rho^2 \hbar \pi}\right)^{1/4} \frac{1}{\sqrt{2^n n!}} \exp\left[\frac{i}{\hbar} P_p^{cl}(t) Q - \frac{1}{4\rho^2 \hbar}\left(Q - Q_p^{cl}(t)\right)^2\right]$$
$$\times H_n\left[\left(\frac{1}{2\rho^2 \hbar}\right)^{1/2} \left(Q - Q_p^{cl}(t)\right)\right], \tag{21.a}$$

$$\lambda_n = n + \frac{1}{2}. \tag{21.b}$$

By inserting Eq. (21.a) into Eq. (20), the phases $\Upsilon_n(t)$ read

$$\Upsilon_n(t) = -\left(n + \frac{1}{2}\right)\frac{t\beta_1}{\rho^2} - \frac{1}{\hbar}\int_0^t \left[\frac{1}{4\beta_1}\left(\dot{Q}_p^{cl}(t')\right)^2 - \beta_2\left(Q_p^{cl}(t')\right)^2 - \frac{\beta_3^2(t')}{4\beta_1}\right]dt', \tag{22}$$

and using Eqs. (19) and (21.a), the solution of Eq. (13) reads

$$\psi_n(Q,t) = e^{i\Upsilon_n(t)} \left(\frac{1}{2\rho^2 \hbar \pi}\right)^{1/4} \frac{1}{\sqrt{2^n n!}} \exp\left[\frac{i}{\hbar} P_p^{cl}(t) Q - \frac{1}{4\rho^2 \hbar}\left(Q - Q_p^{cl}(t)\right)^2\right]$$
$$\times H_n\left[\left(\frac{1}{2\rho^2 \hbar}\right)^{1/2} \left(Q - Q_p^{cl}(t)\right)\right]. \tag{23}$$

Therefore, from the coefficients $\beta_m$ in Eq. (11) and from Eqs. (16), (18), (22) and (23), the solution of the equation for $K(\mathbf{Q}, \mathbf{P}, t)$ is

$$\psi_{n_1,n_2}^{NCPS}(\mathbf{Q},t) = \psi_{n_1}(Q_1,t)\psi_{n_2}(Q_2,t)$$
$$= \exp[i\Upsilon(n_1,n_2,t)] \left(\frac{1}{2\rho_1^2 \hbar \pi}\right)^{1/4} \left(\frac{1}{2\rho_2^2 \hbar \pi}\right)^{1/4} \frac{1}{\sqrt{2^{n_1+n_2} n_1! n_2!}}$$
$$\times \exp\left[\frac{i}{\hbar} Q_1 P_{p,1}^{cl}(t) - \frac{1}{4\rho_1^2 \hbar}\left(Q_1 - Q_{p,1}^{cl}(t)\right)^2\right] H_{n_1}\left[\left(\frac{1}{2\rho_1^2 \hbar}\right)^{1/2} \left(Q_1 - Q_{p,1}^{cl}(t)\right)\right]$$
$$\times \exp\left[\frac{i}{\hbar} Q_2 P_{p,2}^{cl}(t) - \frac{1}{4\rho_2^2 \hbar}\left(Q_2 - Q_{p,2}^{cl}(t)\right)^2\right] H_{n_2}\left[\left(\frac{1}{2\rho_2^2 \hbar}\right)^{1/2} \left(Q_2 - Q_{p,2}^{cl}(t)\right)\right], \tag{24}$$

where

$$\rho_1^2 = \rho_2^2 = \frac{1}{2M\omega_1}, \tag{25.a}$$

$$\ddot{Q}^{cl}_{p,1}(t) + \omega_1^2 Q^{cl}_{p,1}(t) = -\frac{1}{M}\xi_1(t) + \dot{\Omega}_1(t), \tag{25.b}$$

$$\ddot{Q}^{cl}_{p,2}(t) + \omega_1^2 Q^{cl}_{p,2}(t) = -\frac{1}{M}\xi_2(t) + \dot{\Omega}_2(t), \tag{25.c}$$

$$\ddot{P}^{cl}_{p,1}(t) + \omega_1^2 P^{cl}_{p,1}(t) = -M\omega_1^2 \Omega_1(t) - \dot{\xi}_1(t), \tag{25.d}$$

$$\ddot{P}^{cl}_{p,2}(t) + \omega_1^2 P^{cl}_{p,2}(t) = -M\omega_1^2 \Omega_2(t) - \dot{\xi}_2(t), \tag{25.e}$$

$$\Upsilon(n_1, n_2, t) = -(n_1 + n_2 + 1)\omega_1 t$$
$$-\frac{M}{2\hbar}\int_0^t \left[\left(\dot{Q}^{cl}_{p,1}(t')\right)^2 - \omega_1^2 \left(Q^{cl}_{p,1}(t')\right)^2 - \Omega_1^2(t')\right]dt'$$
$$-\frac{M}{2\hbar}\int_0^t \left[\left(\dot{Q}^{cl}_{p,2}(t')\right)^2 - \omega_1^2 \left(Q^{cl}_{p,2}(t')\right)^2 - \Omega_2^2(t')\right]dt'. \tag{25.f}$$

At last, in terms of the original commutative variables, the exact eigenfunctions of the Schrödinger equation (Eq. (7)) are given by

$$\psi^{NCPS}_{n_1,n_2}(x_1, x_2, t) = \left(\frac{1}{2\rho^2 \hbar \pi 2^{n_1+n_2} n_1! n_2!}\right)^{1/2} exp[i\Upsilon(n_1, n_2, t)]$$
$$\times exp\left\{\frac{i}{\hbar}f(t)x_1 - \frac{1}{4\rho^2 \hbar}(x_1 - T(t))^2\right\} exp\left\{\frac{i}{\hbar}g(t)x_2 - \frac{1}{4\rho^2 \hbar}(x_2 - \sigma(t))^2\right\}$$
$$\times H_{n_1}\left[\left(\frac{1}{2\rho^2 \hbar}\right)^{1/2}\left(cos\phi(t)x_1 - sin\phi(t)x_2 - Q^{cl}_{p,1}(t)\right)\right]$$
$$\times H_{n_2}\left[\left(\frac{1}{2\rho^2 \hbar}\right)^{1/2}\left(sin\phi(t)x_1 + cos\phi(t)x_2 - Q^{cl}_{p,2}(t)\right)\right], \tag{26}$$

where

$$\rho_1^2 = \rho_2^2 = \rho^2, \tag{27.a}$$
$$f(t) = P^{cl}_{p,1}(t)cos\phi(t) + P^{cl}_{p,2}(t)sin\phi(t), \tag{27.b}$$
$$T(t) = Q^{cl}_{p,1}(t)cos\phi(t) + Q^{cl}_{p,2}(t)sin\phi(t), \tag{27.c}$$
$$g(t) = -P^{cl}_{p,1}(t)sin\phi(t) + P^{cl}_{p,2}(t)cos\phi(t), \tag{27.d}$$
$$\sigma(t) = -Q^{cl}_{p,1}(t)sin\phi(t) + Q^{cl}_{p,2}(t)cos\phi(t). \tag{27.e}$$

Now, let us consider the simplest case where $n_1 = n_2 = 0$. Thus, $\psi^{NCPS}_{0,0}(x_1, x_2, t)$ and its Fourier Transform are, respectively, given by

$$\psi_{0,0}^{NCPS}(r,t) = \left(\frac{1}{2\rho^2\hbar\pi}\right)^{1/2} exp[i\Upsilon(0,0,t)] \, exp\left\{\frac{i}{\hbar}f(t)x_1 - \frac{1}{4\rho^2\hbar}(x_1 - T(t))^2\right\}$$

$$\times exp\left\{\frac{i}{\hbar}g(t)x_2 - \frac{1}{4\rho^2\hbar}(x_2 - \sigma(t))^2\right\}, \tag{28}$$

$$\Xi_{0,0}^{NCPS}(p,t) = \left(\frac{2\rho^2}{\hbar\pi}\right)^{1/2} exp[i\Upsilon(0,0,t)] \, exp\left\{-\frac{i}{\hbar}T(t)(p_1 - f(t)) - \frac{\rho^2}{\hbar}(p_1 - f(t))^2\right\}$$

$$\times exp\left\{-\frac{i}{\hbar}\sigma(t)(p_2 - g(t)) - \frac{\rho^2}{\hbar}(p_2 - g(t))^2\right\}, \tag{29}$$

In the commutative Quantum Mechanics, the Fisher information for position ($F_r$) and momentum ($F_p$) spaces hold a reciprocity relation with the commutative uncertainties $\Delta r^2$ and $\Delta p^2$. By considering the 2D continuous probability densities $\chi(r,t) = |\psi(r,t)|^2$ and $\vartheta(p,t) = |\Xi(p,t)|^2$, the explicit expressions for $F_r$ and $F_p$ are obtained from the so-called Cramer-Rao inequalities [33, 34]

$$F_r(\Delta r^2) \geq 4, \tag{30.a}$$

$$F_p(\Delta p^2) \geq 4, \tag{30.b}$$

where

$$F_r = \sum_{l=1}^{2} F_{x_l} = \sum_{l=1}^{2} \int \left[\frac{1}{\chi(r,t)}\left(\frac{\partial \chi(r,t)}{\partial x_l}\right)^2\right] d^2r, \tag{31.a}$$

$$F_p = \sum_{l=1}^{2} F_{p_l} = \sum_{l=1}^{2} \int \left[\frac{1}{\vartheta(p,t)}\left(\frac{\partial \vartheta(p,t)}{\partial p_l}\right)^2\right] d^2p. \tag{31.b}$$

By using Eqs.(4), the expressions of $\Delta \hat{r}^2$ and $\Delta \hat{p}^2$ for $n_1 = n_2 = 0$ read

$$\Delta \hat{r}^2 = \Delta r^2 + \frac{\theta^2}{4\hbar^2}\Delta p^2, \tag{32.a}$$

$$\Delta \hat{p}^2 = \Delta p^2 + \frac{\eta^2}{4\hbar^2}\Delta r^2, \tag{32.b}$$

indicating that $\Delta \hat{r}^2$ and $\Delta \hat{p}^2$ are both given by combinations of $\Delta r$ and $\Delta p$. Therefore, to have a noncommutative version of the Cramer-Rao inequalities in the form

$$F_{\hat{r}}(\Delta \hat{r}^2) \geq 4, \qquad (33.a)$$

$$F_{\hat{p}}(\Delta \hat{p}^2) \geq 4, \qquad (33.b)$$

the expressions for $F_{\hat{r}}$ and $F_{\hat{p}}$ should be given by

$$F_{\hat{r}} = \frac{F_r}{1 + \frac{\theta^2}{4\hbar^2}\frac{F_r}{F_p}}, \qquad (34.a)$$

$$F_{\hat{p}} = \frac{F_p}{1 + \frac{\eta^2}{4\hbar^2}\frac{F_p}{F_r}}. \qquad (34.b)$$

From Eqs. (8.a), (8.b), (25.a), (27.a), (28), (29), (31) and (34), the expressions for $F_{\hat{r}}$ and $F_{\hat{p}}$ for the harmonic oscillator in a time-dependent electric field on a 2D noncommutative background read

$$F_{\hat{r}}(\theta, \eta) = \frac{4m\omega_0}{\hbar}\left(1 + \frac{m^2\omega_0^2\theta^2}{4\hbar^2}\right)^{-1/2}\left(1 + \frac{\eta^2}{4\hbar^2 m^2\omega_0^2}\right)^{1/2}$$

$$\times \left[1 + \frac{m^2\omega_0^2\theta^2}{4\hbar^2}\left(1 + \frac{m^2\omega_0^2\theta^2}{4\hbar^2}\right)^{-1}\left(1 + \frac{\eta^2}{4\hbar^2 m^2\omega_0^2}\right)\right]^{-1}, (35.a)$$

$$F_{\hat{p}}(\theta, \eta) = \frac{4}{\hbar m\omega_0}\left(1 + \frac{m^2\omega_0^2\theta^2}{4\hbar^2}\right)^{1/2}\left(1 + \frac{\eta^2}{4\hbar^2 m^2\omega_0^2}\right)^{-1/2}$$

$$\times \left[1 + \frac{\eta^2}{4\hbar^2 m^2\omega_0^2}\left(1 + \frac{m^2\omega_0^2\theta^2}{4\hbar^2}\right)\left(1 + \frac{\eta^2}{4\hbar^2 m^2\omega_0^2}\right)^{-1}\right]^{-1}, (35.b)$$

which are time-independent as well as $\Delta \hat{r}$ and $\Delta \hat{p}$. These results agree with those found by Choi [32] who investigated the system described by a general Hamiltonian $H = A(t)p^2 + B(t)(xp + px) + C(t)x^2 + D(t)x + E(t)p + F(t)$, where $[x, p] = i\hbar$.

On the other hand, in a 2D commutative quantum mechanics the Shannon entropy for position ($S_r$) and momentum ($S_p$) spaces is defined as [35]

$$S_r = -\int \chi(\mathbf{r},t) \ln[\chi(\mathbf{r},t)]\, d^2\mathbf{r}, \qquad (36.a)$$

$$S_p = -\int \vartheta(\boldsymbol{p},t) \ln[\vartheta(\boldsymbol{p},t)]\, d^2\boldsymbol{p}, \qquad (36.b)$$

and represent the sum over the probability densities associated to the observables $\boldsymbol{r}$ and $\boldsymbol{p}$, respectively. Thus, a direct extension to noncommutative space allows one to rewrite Eqs. (36) by simply replacing $\boldsymbol{r}$ and $\boldsymbol{p}$ by $\hat{\boldsymbol{r}}$ and $\hat{\boldsymbol{p}}$, respectively. From Eqs. (8.a), (8.b), (25.a), (27.a), (28), (29) and (36), the expression for $S_{\hat{r}}$ and $S_{\hat{p}}$ read

$$S_{\hat{r}}(\theta,\eta) = 1 + \ln\pi + \ln\left[\frac{\hbar}{m\omega_0}\left(1+\frac{m^2\omega_0^2\theta^2}{4\hbar^2}\right)^{1/2}\left(1+\frac{\eta^2}{4\hbar^2 m^2 \omega_0^2}\right)^{-1/2}\right], (37.a)$$

$$S_{\hat{p}}(\theta,\eta) = 1 + \ln\pi - \ln\left[\frac{1}{\hbar m\omega_0}\left(1+\frac{m^2\omega_0^2\theta^2}{4\hbar^2}\right)^{\frac{1}{2}}\left(1+\frac{\eta^2}{4\hbar^2 m^2 \omega_0^2}\right)^{-\frac{1}{2}}\right], (37.b)$$

which are time-independent and satisfy the Bialynicki-Birula-Mycielski entropic uncertainty relation [36]

$$S_{\hat{r}}(\theta,\eta) + S_{\hat{p}}(\theta,\eta) = 2(1 + \ln\pi + \ln\hbar). \qquad (38)$$

Plots of $F_{\hat{r}}(\theta,\eta)$, $F_{\hat{p}}(\theta,\eta)$, $S_{\hat{r}}(\theta,\eta)$ and $S_{\hat{p}}(\theta,\eta)$ are shown in Figs. 1 (a)-(d), respectively. We observe that for a fixed value of $\eta$, $F_{\hat{r}}(\theta,\eta)$ ($S_{\hat{r}}(\theta,\eta)$) decreases (increases) with increasing $\theta$, indicating that the accuracy in predicting the localization of the particle is decreasing. On the other hand, for a fixed value of $\eta$, $F_{\hat{p}}(\theta,\eta)$ ($S_{\hat{p}}(\theta,\eta)$) increases (decreases) with increasing $\theta$, indicating that the accuracy in predicting momentum increases. We can understand these results by analyzing the uncertainty relation for the noncommutative operators $\hat{x}_1$ and $\hat{x}_2$

$$\Delta\hat{x}_1 \Delta\hat{x}_2 \geq \frac{\theta}{2}. \qquad (39)$$

From Eqs. (4.a) and (28) one obtains $\Delta\hat{x}_1 = \Delta\hat{x}_2$, and inequality (39) reads

$$\Delta\hat{r} \geq \sqrt{\theta}, \qquad (40)$$

indicating that noncommutativity between spatial-spatial coordinates decreases the accuracy in measuring the particle's position as previously observed by Snyder [37] in investigating noncommutative space-time structure.

We also observed that, for a fixed value of $\theta$, $F_{\hat{r}}(\theta,\eta)$ ($S_{\hat{r}}(\theta,\eta)$) increases (decreases) with increasing $\eta$, indicating that the accuracy in predicting the localization of the particle increases. On the other hand, for a fixed value of $\theta$, $F_{\hat{p}}(\theta,\eta)$ ($S_{\hat{p}}(\theta,\eta)$) decreases (increases) with increasing $\eta$, indicating that the accuracy in predicting momentum decreases. These results become clear when we analyze the uncertainty relation for the noncommutative observables $\hat{p}_1$ and $\hat{p}_2$ given by

$$\Delta\hat{p}_1 \Delta\hat{p}_2 \geq \frac{\eta}{2}. \tag{41}$$

Since $\Delta\hat{p}_1 = \Delta\hat{p}_2$, we obtain from Eqs. (4.b) and (28) that

$$\Delta\hat{\boldsymbol{p}} \geq \sqrt{\eta}, \tag{42}$$

which indicates that the accuracy in measuring $\Delta\hat{\boldsymbol{p}}$ decreases with increasing $\eta$.

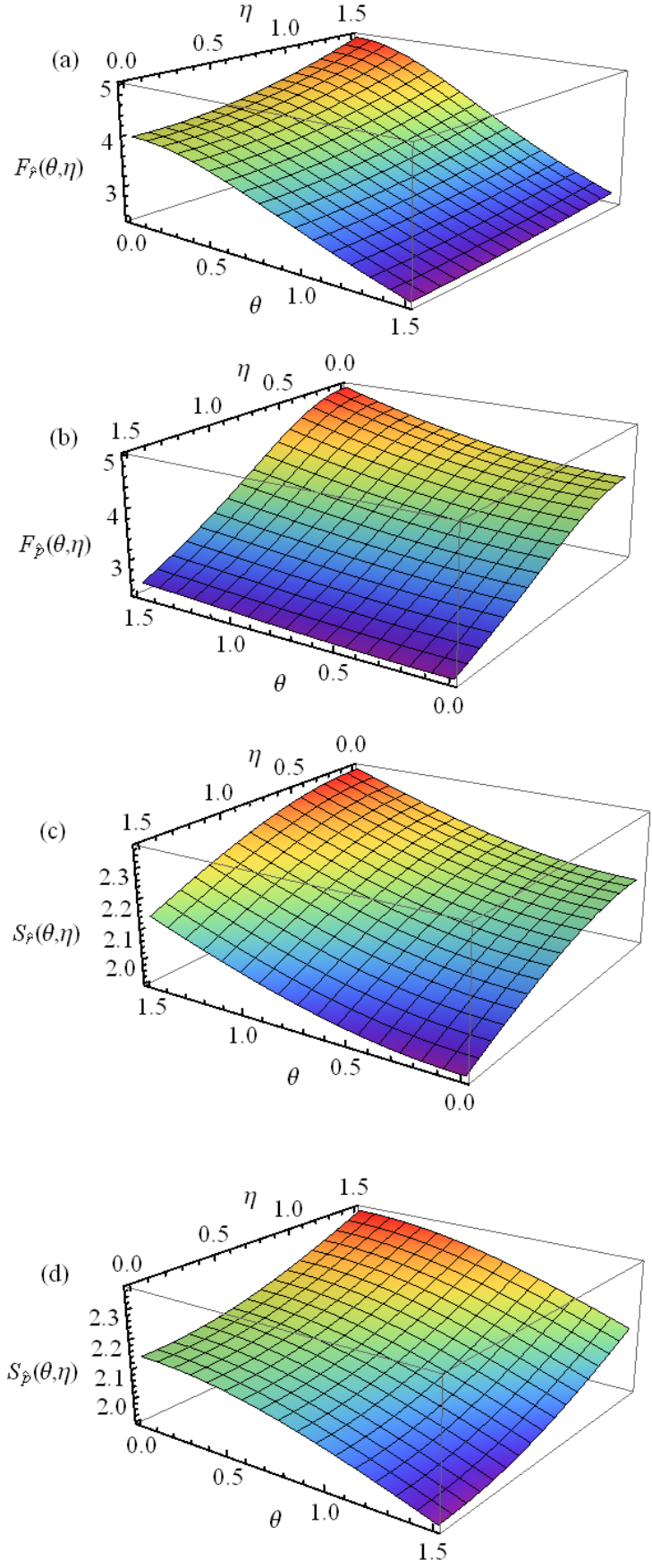

**Figure 1**. Plots of (a) $F_{\hat{r}}(\theta,\eta)$, (b) $F_{\hat{p}}(\theta,\eta)$, (c) $S_{\hat{r}}(\theta,\eta)$ and (d) $S_{\hat{p}}(\theta,\eta)$ for a harmonic oscillator in a time-depent electric field on a 2D noncommutative background. In the figures we used $m = \omega_0 = \hbar = 1$.

## 3. Harmonic oscillator in a time-dependent electric field on a 3D noncommutative phase space.

In 3D the Eqs. (1) and (2) remain valid for $m, n = 1, 2, 3$. By considering $\theta_{mn} = \epsilon^{mnk}\theta_k$ and $\eta_{mn} = \epsilon^{mnk}\eta_k$, with $\boldsymbol{\theta} = (0, 0, \theta)$ and $\boldsymbol{\eta} = (0, 0, \eta)$, we obtain the following tranformations between commutative and noncommutative variables

$$\hat{x}_1 = x_1 - \frac{1}{2\hbar}\theta p_2, \qquad \hat{x}_2 = x_2 + \frac{1}{2\hbar}\theta p_1, \qquad \hat{x}_3 = x_3, \qquad (43.a)$$

$$\hat{p}_1 = p_1 + \frac{1}{2\hbar}\eta x_2, \qquad \hat{p}_2 = p_2 - \frac{1}{2\hbar}\eta x_1, \qquad \hat{p}_3 = p_3, \qquad (43.b)$$

where $\hat{x}_3 (\hat{p}_3)$ commutes with $\hat{x}_1 (\hat{p}_1)$ and $\hat{x}_2 (\hat{p}_2)$.

Again, to calculate the Fisher information and Shannon entropy for a harmonic oscillator in a time-dependent electric field in a 3D noncommutative background, we have to solve the time-dependent Schrödinger equation

$$i\hbar \partial_t \psi^{NCPS}(\hat{\boldsymbol{r}}, t) = H^{NCPS}(\hat{\boldsymbol{r}}, \hat{\boldsymbol{p}}, t)\psi^{NCPS}(\hat{\boldsymbol{r}}, t), \qquad (44)$$

where $H^{NCPS}$ is given by

$$H^{NCPS}(\hat{\boldsymbol{r}}, \hat{\boldsymbol{p}}, t) = \frac{1}{2m}(\hat{p}_1^2 + \hat{p}_2^2 + \hat{p}_3^2) + \frac{1}{2}m\omega_0^2(\hat{x}_1^2 + \hat{x}_2^2 + \hat{x}_3^2) + q\Phi(\hat{\boldsymbol{r}}, t), (45)$$

and $\Phi(\hat{\boldsymbol{r}}, t) = E_1(t)\hat{x}_1 + E_2(t)\hat{x}_2 + E_3(t)\hat{x}_3$ is the eletrostatic potential.

By applying the Bopp shift method to Eq. (44) we obtain

$$i\hbar\partial_t \psi^{NCPS} = \left[\frac{1}{2M}(p_1^2 + p_2^2) + \frac{1}{2m}p_3^2 + \frac{1}{2}M\omega_1^2(x_1^2 + x_2^2) + \frac{1}{2}m\omega_0^2 x_3^2 - \omega_2 L_3\right]\psi^{NCPS}$$
$$+[A(t)p_1 + B(t)p_2 + C(t)x_1 + D(t)x_2 + F(t)x_3]\psi^{NCPS}(\boldsymbol{r}, t), \quad (46)$$

where $M, \omega_1^2, \omega_2, A(t), B(t), C(t), D(t)$ and $L_3$ are given by Eqs. (8) and $F(t) = qE_3(t)$.

To find $\psi^{NCPS}(\boldsymbol{r}, t)$ let us first consider the time-dependent canonical transformation whose generating function is

$$G(x_1, x_2, x_3, P_1, P_2, P_3, t) = (P_1 cos\phi + P_2 sin\phi)x_1 + (-P_1 sin\phi + P_2 cos\phi)x_2$$
$$+ x_3 P_3, \qquad (47)$$

where $\phi = \phi(t) = \omega_2 t$.

From Eq. (47) we obtain the new canonical variables

$$p_1 = \frac{\partial G}{\partial x_1} = P_1 cos\phi + P_2 sin\phi, \qquad (48.a)$$

$$p_2 = \frac{\partial G}{\partial x_2} = -P_1 sin\phi + P_2 cos\phi, \qquad (48.b)$$

$$p_3 = \frac{\partial G}{\partial x_3} = P_3, \qquad (48.c)$$

$$Q_1 = \frac{\partial G}{\partial P_1} = cos\phi x_1 - sin\phi x_2, \qquad (48.d)$$

$$Q_2 = \frac{\partial G}{\partial P_2} = sin\phi x_1 + cos\phi x_2, \qquad (48.e)$$

$$Q_3 = \frac{\partial G}{\partial P_3} = x_3, \qquad (48.f)$$

and the new Hamiltonian, $K(Q_1, Q_2, Q_3, P_1, P_2, P_3, t) = H^{NCPS}(x_1, x_2, x_3, p_1, p_2, p_3, t) + \partial_t G$, reads

$$K(\boldsymbol{Q}, \boldsymbol{P}, t) = \frac{1}{2M} P_1^2 + + \frac{1}{2} M\omega_1^2 Q_1^2 + \Omega_1(t) P_1 + \xi_1(t) Q_1$$
$$+ \frac{1}{2M} P_2^2 + + \frac{1}{2} M\omega_1^2 Q_2^2 + \Omega_2(t) P_2 + \xi_2(t) Q_2$$
$$+ \frac{1}{2m} P_3^2 + \frac{1}{2} m\omega_0^2 Q_3^2 + F(t) Q_3, \qquad (49)$$

where $\Omega_1(t)$, $\Omega_2(t)$, $\xi_1(t)$ and $\xi_2(t)$ are given by Eqs. (12).

Since Eq. (49) represents the sum of three uncoupled Hamiltonians of time-dependent quadratic harmonic oscillators, the procedure performed in Section 2 can be straighforwardly applied and we find that the exact solution of the Schrödinger equation for $K(\boldsymbol{Q}, \boldsymbol{P}, t)$ is

$$\psi_{n_1,n_2,n_3}^{NCPS}(\mathbf{Q},t) = \psi_{n_1}(Q_1,t)\psi_{n_2}(Q_2,t)\psi_{n_3}(Q_3,t)$$
$$= exp[i\Upsilon(n_1,n_2,n_3,t)]\varphi_{n_1}(Q_1,t)\varphi_{n_2}(Q_2,t)\varphi_{n_3}(Q_3,t)$$
$$= exp[i\Upsilon(n_1,n_2,n_3,t)]\left(\frac{1}{2^3\rho_1^2\rho_2^2\rho_3^2\hbar^3\pi^3}\right)^{1/4}\frac{1}{\sqrt{2^{n_1+n_2+n_3}\,n_1!\,n_2!\,n_3!}}$$
$$\times exp\left[\frac{i}{\hbar}Q_1 P_{p,1}^{cl}(t) - \frac{1}{4\rho_1^2\hbar}\left(Q_1 - Q_{p,1}^{cl}(t)\right)^2\right]H_{n_1}\left[\left(\frac{1}{2\rho_1^2\hbar}\right)^{1/2}\left(Q_1 - Q_{p,1}^{cl}(t)\right)\right]$$
$$\times exp\left[\frac{i}{\hbar}Q_2 P_{p,2}^{cl}(t) - \frac{1}{4\rho_2^2\hbar}\left(Q_2 - Q_{p,2}^{cl}(t)\right)^2\right]H_{n_2}\left[\left(\frac{1}{2\rho_2^2\hbar}\right)^{1/2}\left(Q_2 - Q_{p,2}^{cl}(t)\right)\right]$$
$$\times exp\left[\frac{i}{\hbar}Q_3 P_{p,3}^{cl}(t) - \frac{1}{4\rho_3^2\hbar}\left(Q_3 - Q_{p,3}^{cl}(t)\right)^2\right]H_{n_3}\left[\left(\frac{1}{2\rho_3^2\hbar}\right)^{1/2}\left(Q_3 - Q_{p,3}^{cl}(t)\right)\right], \quad (50)$$

where $\rho_1^2$, $\rho_2^2$, $Q_{p,1}^{cl}(t)$, $Q_{p,2}^{cl}(t)$, $P_{p,1}^{cl}(t)$ and $P_{p,2}^{cl}(t)$ are given by Eqs. (25.(a)-(e)) and $\rho_3^2$, $Q_{p,3}^{cl}(t)$, $P_{p,3}^{cl}(t)$ and $\Upsilon(n_1,n_2,n_3,t)$ satisfy the equations

$$\rho_3^2 = \frac{1}{2m\omega_0}, \quad (51.a)$$

$$\ddot{Q}_{p,3}^{cl}(t) + \omega_0^2 Q_{p,3}^{cl}(t) = -\frac{1}{m}F(t), \quad (51.b)$$

$$\ddot{P}_{p,3}^{cl}(t) + \omega_0^2 P_{p,3}^{cl}(t) = -\dot{F}(t), \quad (51.c)$$

$$\Upsilon(n_1,n_2,n_3,t) = -(n_1+n_2+1)\omega_1 t - \left(n_3+\frac{1}{2}\right)\omega_0 t$$
$$-\frac{M}{2\hbar}\int_0^t\left[\left(\dot{Q}_{p,1}^{cl}(t')\right)^2 - \omega_1^2\left(Q_{p,1}^{cl}(t')\right)^2 - \Omega_1^2(t')\right]dt'$$
$$-\frac{M}{2\hbar}\int_0^t\left[\left(\dot{Q}_{p,2}^{cl}(t')\right)^2 - \omega_1^2\left(Q_{p,2}^{cl}(t')\right)^2 - \Omega_2^2(t')\right]dt'$$
$$-\frac{m}{2\hbar}\int_0^t\left[\left(\dot{Q}_{p,3}^{cl}(t')\right)^2 - \omega_0^2\left(Q_{p,3}^{cl}(t')\right)^2\right]dt'. \quad (51.d)$$

In terms of the original variables, the exact eigenfunctions of the Schrödinger equation (44) are given by

$$\psi_{n_1,n_2,n_3}^{NCPS}(\mathbf{r},t) = \left(\frac{1}{2\rho^2\hbar\pi 2^{n_1+n_2+n_3}\,n_1!\,n_2!\,n_3!}\right)^{1/2}\left(\frac{1}{2\rho_3^2\hbar\pi}\right)^{1/4}exp[i\Upsilon(n_1,n_2,n_3,t)]$$
$$\times exp\left\{\frac{i}{\hbar}fx_1 - \frac{1}{4\rho^2\hbar}(x_1-T)^2\right\}exp\left\{\frac{i}{\hbar}gx_2 - \frac{1}{4\rho^2\hbar}(x_2-\sigma)^2\right\}$$

$$\times exp\left\{\frac{i}{\hbar}P_{p,3}^{cl}x_3 - \frac{1}{4\rho_3^2\hbar}\left(x_3 - Q_{p,3}^{cl}(t)\right)^2\right\}$$

$$\times H_{n_1}\left[\left(\frac{1}{2\rho^2\hbar}\right)^{1/2}\left(cos\phi x_1 - sin\phi x_2 - Q_{p,1}^{cl}(t)\right)\right]$$

$$\times H_{n_2}\left[\left(\frac{1}{2\rho^2\hbar}\right)^{1/2}\left(sin\phi x_1 + cos\phi x_2 - Q_{p,2}^{cl}(t)\right)\right]$$

$$\times H_{n_3}\left[\left(\frac{1}{2\rho_3^2\hbar}\right)^{1/2}\left(x_3 - Q_{p,3}^{cl}(t)\right)\right], \qquad (52)$$

with $\rho^2, f(t), T(t), g(t)$ and $\sigma(t)$ given by Eqs. (27).

Consider the simplest case where $n_1 = n_2 = n_3 = 0$. Thus, $\psi_{0,0,0}^{NCPS}(r,t)$ and its Fourier Transform are, respectively, given by

$$\psi_{0,0,0}^{NCPS}(r,t) = \left(\frac{1}{2\rho^2\hbar\pi}\right)^{1/2}\left(\frac{1}{2\rho_3^2\hbar\pi}\right)^{1/4} exp[i\Upsilon(0,0,0,t)]$$

$$\times exp\left\{\frac{i}{\hbar}fx_1 - \frac{1}{4\rho^2\hbar}(x_1 - T)^2 + \frac{i}{\hbar}gx_2 - \frac{1}{4\rho^2\hbar}(x_2 - \sigma)^2\right\}$$

$$\times exp\left\{\frac{i}{\hbar}P_{p,3}^{cl}x_3 - \frac{1}{4\rho_3^2\hbar}\left(x_3 - Q_{p,3}^{cl}(t)\right)^2\right\}, \qquad (53)$$

$$\Xi_{0,0,0}^{NCPS}(p,t) = \left(\frac{2\rho^2}{\hbar\pi}\right)^{1/2}\left(\frac{2\rho_3^2}{\hbar\pi}\right)^{1/4} exp[i\Upsilon(0,0,0,t)]$$

$$\times exp\left\{-\frac{i}{\hbar}T(p_1 - f) - \frac{\rho^2}{\hbar}(p_1 - f)^2 - \frac{i}{\hbar}\sigma(p_2 - g) - \frac{\rho^2}{\hbar}(p_2 - g)^2\right\}$$

$$\times exp\left\{-\frac{i}{\hbar}Q_{p,3}^{cl}\left(p_3 - P_{p,3}^{cl}\right) - \frac{\rho_3^2}{\hbar}\left(p_3 - P_{p,3}^{cl}\right)^2\right\}. \qquad (54)$$

As in the 2D case, we need to construct a noncommutative version of the Cramer-Rao inequalities to define the noncommutative expressions of the Fisher informations. The 3D version of the Cramer-Rao inequalities in a commutative space read [33, 34]

$$F_r(\Delta r^2) \geq 9, \qquad (55.a)$$
$$F_p(\Delta p^2) \geq 9, \qquad (55.b)$$

where $F_r$ and $F_p$ are given by Eqs. (31) with the summation extending from 1 to 3 ($l = 1, 2, 3$).

The uncertainties calculated in the ground state (Eq. 53) on the noncommutative 3D space read

$$\Delta \hat{r}^2 = \Delta r^2 + \frac{\theta^2}{4\hbar^2}(\Delta p_1^2 + \Delta p_2^2), \tag{56.a}$$

$$\Delta \hat{p}^2 = \Delta p^2 + \frac{\eta^2}{4\hbar^2}(\Delta x_1^2 + \Delta x_2^2). \tag{56.b}$$

The noncommutative version of the Cramer-Rao inequalities in the form

$$F_{\hat{r}}(\Delta \hat{r}^2) \geq 9, \tag{57.a}$$

$$F_{\hat{p}}(\Delta \hat{p}^2) \geq 9, \tag{57.b}$$

is obtained if the expressions for $F_{\hat{r}}$ and $F_{\hat{p}}$ are given by

$$F_{\hat{r}} = \frac{F_r}{1 + \frac{\theta^2}{9\hbar^2} \frac{F_r}{(F_{p_1} + F_{p_2})}}, \tag{58.a}$$

$$F_{\hat{p}} = \frac{F_p}{1 + \frac{\eta^2}{9\hbar^2} \frac{F_p}{(F_{x_1} + F_{x_2})}}. \tag{58.b}$$

From Eqs. (8.a), (8.b), (25.a), (27.a), (51.a), (53), (54) and (58), $F_{\hat{r}}$ and $F_{\hat{p}}$ are given by

$$F_{\hat{r}}(\theta, \eta) = \frac{4m\omega_0}{\hbar} \left[ \left(1 + \frac{m^2\omega_0^2\theta^2}{4\hbar^2}\right)^{-1/2} \left(1 + \frac{\eta^2}{4\hbar^2 m^2\omega_0^2}\right)^{1/2} + \frac{1}{2} \right]$$

$$\times \left\{ 1 + \frac{m^2\omega_0^2\theta^2}{9\hbar^2} \left(1 + \frac{m^2\omega_0^2\theta^2}{4\hbar^2}\right)^{-1/2} \left(1 + \frac{\eta^2}{4\hbar^2 m^2\omega_0^2}\right)^{1/2} \right.$$

$$\left. \times \left[ \left(1 + \frac{m^2\omega_0^2\theta^2}{4\hbar^2}\right)^{-1/2} \left(1 + \frac{\eta^2}{4\hbar^2 m^2\omega_0^2}\right)^{1/2} + \frac{1}{2} \right] \right\}^{-1} \tag{59.a}$$

$$F_{\hat{p}}(\theta,\eta) = \frac{4}{\hbar m\omega_0}\left[\left(1+\frac{m^2\omega_0^2\theta^2}{4\hbar^2}\right)^{1/2}\left(1+\frac{\eta^2}{4\hbar^2 m^2\omega_0^2}\right)^{-1/2}+\frac{1}{2}\right]$$
$$\times\left\{1+\frac{\eta^2}{9\hbar^2 m^2\omega_0^2}\left(1+\frac{m^2\omega_0^2\theta^2}{4\hbar^2}\right)^{1/2}\left(1+\frac{\eta^2}{4\hbar^2 m^2\omega_0^2}\right)^{-1/2}\right.$$
$$\left.\times\left[\left(1+\frac{m^2\omega_0^2\theta^2}{4\hbar^2}\right)^{1/2}\left(1+\frac{\eta^2}{4\hbar^2 m^2\omega_0^2}\right)^{-1/2}+\frac{1}{2}\right]^{-1}\right\} \quad (59.b)$$

As discussed before, there is no problem to define the respective Shannon entropy on position and momentum in the 3D noncommutative space. By using Eqs. (8.a), (8.b), (25.a), (27.a), (36), (51.a), (53) and (54), the expression of $S_{\hat{r}}$ and $S_{\hat{p}}$ read

$$S_{\hat{r}}(\theta,\eta) = \frac{3}{2}+\frac{3}{2}\ln\pi$$
$$+\ln\left[\left(\frac{\hbar}{m\omega_0}\right)^{3/2}\left(1+\frac{m^2\omega_0^2\theta^2}{4\hbar^2}\right)^{1/2}\left(1+\frac{\eta^2}{4\hbar^2 m^2\omega_0^2}\right)^{-1/2}\right], \quad (60.a)$$

$$S_{\hat{p}}(\theta,\eta) = \frac{3}{2}+\frac{3}{2}\ln\pi$$
$$-\ln\left[\left(\frac{1}{\hbar m\omega_0}\right)^{3/2}\left(1+\frac{m^2\omega_0^2\theta^2}{4\hbar^2}\right)^{1/2}\left(1+\frac{\eta^2}{4\hbar^2 m^2\omega_0^2}\right)^{-1/2}\right], \quad (60.b)$$

which satisfy the following Bialynicki-Birula-Mycielski entropic uncertainty relation [36]

$$S_{\hat{r}}(\theta,\eta) + S_{\hat{p}}(\theta,\eta) = 3(1+\ln\pi+\ln\hbar). \quad (61)$$

Plots of $F_{\hat{r}}(\theta,\eta)$, $F_{\hat{p}}(\theta,\eta)$, $S_{\hat{r}}(\theta,\eta)$ and $S_{\hat{p}}(\theta,\eta)$ are shown in Figs. 2 (a)-(d), respectively. The behaviors observed indicate that the motion along the z direction does not change the overall results obtained in the 2D case.

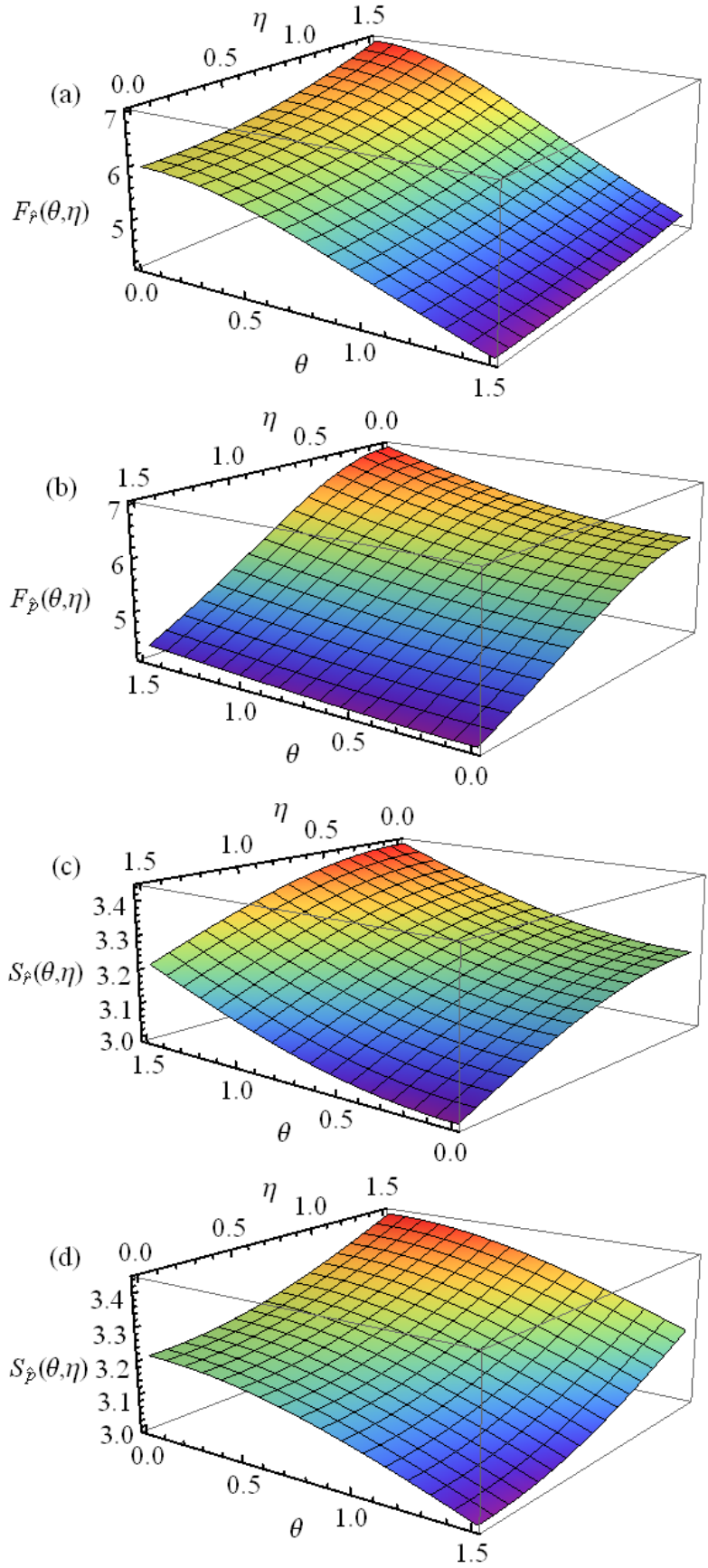

**Figure 2**. Plots of (a) $F_{\hat{r}}(\theta,\eta)$, (b) $F_{\hat{p}}(\theta,\eta)$, (c) $S_{\hat{r}}(\theta,\eta)$ and (d) $S_{\hat{p}}(\theta,\eta)$ for the harmonic oscillator in a time-depent electric field on a 3D noncommutative background. In the figures we used $m = \omega_0 = \hbar = 1$.

## 4. Concluding remarks

In this paper, we used a canonical transformation and the Lewis and Riesenfeld invariant method to obtain the Schrödinger wave function of a harmonic oscillator in a time-varying electric field in 2D and 3D noncommutative spaces. By using appropriate canonical transformations we rewrite the Hamiltonians given by Eqs. (7) and (46) as those given by Eqs. (11) and (49), which correspond to the sum of two and three uncoupled Hamiltonians of harmonic oscillators with linear terms in momentum and position, respectively. The Lewis and Riesenfeld method was therefore used to obtain the wave functions given by Eqs. (26) and (52).

From the wave functions obtained for the lowest lying states (Eqs. (28) and (53)), we calculated the Fisher Information ($F_{\hat{r},\hat{p}}$) and the Shannon Entropy ($S_{\hat{r},\hat{p}}$) for each systems. All the quantities are time-independent. In both cases, we observed that, for a fixed value of $\eta$, $F_{\hat{r}}(\theta,\eta)$ ($S_{\hat{r}}(\theta,\eta)$) decreases (increases) while $F_{\hat{p}}(\theta,\eta)$ ($S_{\hat{p}}(\theta,\eta)$) increases (decreases) with increasing $\theta$. On the other hand, for a fixed value of $\theta$, $F_{\hat{r}}(\theta,\eta)$ ($S_{\hat{r}}(\theta,\eta)$) increases (decreases) while $F_{\hat{p}}(\theta,\eta)$ ($S_{\hat{p}}(\theta,\eta)$) decreases (increases) with increasing $\eta$. These results are consequence of the minimum value of the uncertainties $\Delta\hat{r}$ and $\Delta\hat{p}$, which is controlled by the noncommutative parameters $\theta$ and $\eta$. We also verified that the Bialynicki-Birula-Mycielski inequality and the relation $S_{\hat{r}} + S_{\hat{p}} = D(1 + \ln\pi + \ln\hbar)$, where $D$ is the space dimension, hold.

## Acknowledgments

The authors are grateful to the National Counsel of Scientific and Technological Development (CNPq) of Brazil for financial support.